\documentclass[journal]{IEEEtran}
\usepackage{cite}
\ifCLASSINFOpdf
   \usepackage[pdftex]{graphicx}
\else
\fi
\usepackage{amsmath}
\usepackage{array}
\usepackage{float} 
\usepackage{multirow}
\begin{document}
%
% paper title
% Titles are generally capitalized except for words such as a, an, and, as,
% at, but, by, for, in, nor, of, on, or, the, to and up, which are usually
% not capitalized unless they are the first or last word of the title.
% Linebreaks \\ can be used within to get better formatting as desired.
% Do not put math or special symbols in the title.
\title{Articulatory-WaveNet: Autoregressive Model For
Acoustic-to-Articulatory Inversion}
%
%
% author names and IEEE memberships
% note positions of commas and nonbreaking spaces ( ~ ) LaTeX will not break
% a structure at a ~ so this keeps an author's name from being broken across
% two lines.
% use \thanks{} to gain access to the first footnote area
% a separate \thanks must be used for each paragraph as LaTeX2e's \thanks
% was not built to handle multiple paragraphs
%
\author{Narjes~Bozorg
        and~Michael~Johnson
        % <-this % stops a space
\thanks{N. Bozorg and M. Johnson are with the Department
of Electrical and Computer Engineering, University of Kentucky, Lexington,
KY, 40506 USA (e-mail: narjes.bozorg@uky.edu; mike.johnson@uky.edu).}% <-this % stops a space
\thanks{Manuscript received March 30, 2020.}}
% note the % following the last \IEEEmembership and also \thanks - 
% these prevent an unwanted space from occurring between the last author name
% and the end of the author line. i.e., if you had this:
% 
% \author{....lastname \thanks{...} \thanks{...} }
%                     ^------------^------------^----Do not want these spaces!
%
% a space would be appended to the last name and could cause every name on that
% line to be shifted left slightly. This is one of those "LaTeX things". For
% instance, "\textbf{A} \textbf{B}" will typeset as "A B" not "AB". To get
% "AB" then you have to do: "\textbf{A}\textbf{B}"
% \thanks is no different in this regard, so shield the last } of each \thanks
% that ends a line with a % and do not let a space in before the next \thanks.
% Spaces after \IEEEmembership other than the last one are OK (and needed) as
% you are supposed to have spaces between the names. For what it is worth,
% this is a minor point as most people would not even notice if the said evil
% space somehow managed to creep in.
% The paper headers
\markboth{}%
{Shell \MakeLowercase{\textit{et al.}}: Bare Demo of IEEEtran.cls for IEEE Journals}
% The only time the second header will appear is for the odd numbered pages
% after the title page when using the twoside option.
% 
% *** Note that you probably will NOT want to include the author's ***
% *** name in the headers of peer review papers.                   ***
% You can use \ifCLASSOPTIONpeerreview for conditional compilation here if
% you desire.
% If you want to put a publisher's ID mark on the page you can do it like
% this:
%\IEEEpubid{0000--0000/00\$00.00~\copyright~2015 IEEE}
% Remember, if you use this you must call \IEEEpubidadjcol in the second
% column for its text to clear the IEEEpubid mark.
% use for special paper notices
%\IEEEspecialpapernotice{(Invited Paper)}
% make the title area
\maketitle
% As a general rule, do not put math, special symbols or citations
% in the abstract or keywords.
\begin{abstract}
This paper presents Articulatory-WaveNet, a new approach for acoustic-to-articulator inversion. The proposed system uses the WaveNet speech synthesis architecture, with dilated causal convolutional layers using previous values of the predicted articulatory trajectories conditioned on acoustic features. The system was trained and evaluated on the ElectroMagnetic Articulography corpus of Mandarin Accented English (EMA-MAE),  consisting of 39 speakers including both native English speakers and native Mandarin speakers speaking English. Results show significant improvement in both correlation and RMSE between the generated and true articulatory trajectories for the new method, with an average correlation of 0.83, representing a 36\% relative improvement over the 0.61 correlation obtained with a baseline Hidden Markov Model (HMM)-Gaussian Mixture Model (GMM) inversion framework. To the best of our knowledge, this paper presents the first application of a point-by-point waveform synthesis approach to the problem of acoustic-to-articulatory inversion and the results show improved performance compared to previous methods for speaker dependent acoustic to articulatory inversion.\end{abstract}
% Note that keywords are not normally used for peerreview papers.
% For peer review papers, you can put extra information on the cover
% page as needed:
% \ifCLASSOPTIONpeerreview
% \begin{center} \bfseries EDICS Category: 3-BBND \end{center}
% \fi
%
% For peerreview papers, this IEEEtran command inserts a page break and
% creates the second title. It will be ignored for other modes.
\IEEEpeerreviewmaketitle
\section{Introduction}
% The very first letter is a 2 line initial drop letter followed
% by the rest of the first word in caps.
% 
% form to use if the first word consists of a single letter:
% \IEEEPARstart{A}{demo} file is ....
% 
% form to use if you need the single drop letter followed by
% normal text (unknown if ever used by the IEEE):
% \IEEEPARstart{A}{}demo file is ....
% 
% Some journals put the first two words in caps:
% \IEEEPARstart{T}{his demo} file is ....
% 
% Here we have the typical use of a "T" for an initial drop letter
% and "HIS" in caps to complete the first word.
\IEEEPARstart{S}{peech} production is a highly complex task involving synchronized motor control of more than 100 different muscles. The study of articulatory patterns plays an important role in many fields of study related to speech and signal processing, speech audiology and pathology, and language understanding \cite{murdoch2009acquired,rosenbaum2009human}. The problem of estimating articulatory trajectories from an acoustic signal is known as Acoustic-to-Articulatory Inversion (A2AI), and is applicable to many different domains such as audio-visual synthesis \cite{inproceedings,7506035}, Computer-Aided Language Learning (CALL) \cite{7506035,ji2014speaker,jones2017development}, and Computer Aided Pronunciation Training (CAPT) \cite{7506035,ji2014speaker,jones2017development,li2019improving}.
 
A2AI is a highly nonlinear and non-unique mapping \cite{fant1960acoustic,qin2007empirical}, since different combinations of articulatory movements can result in the same acoustic output \cite{ghosh2010generalized}. Traditional methods for speaker dependent A2AI  include codebook approaches \cite{atal1978inversion,ouni2005modeling}, Kalman filtering \cite{dusan2000acoustic}, Gaussian Mixture Model (GMM) \cite{toda2008statistical} and Hidden Markov Model (HMM) \cite{zhang2008acoustic}. Recently there has been significant progress on A2AI, with several new approaches based on deep learning published in the last few years that have improved the state of the art. Along similar lines, this paper is working toward improving results of our previous GMM-HMM approach to A2AI \cite{bozorg2018comparing} with a new deep learning strategy.
The approach introduced here is based on adapting a waveform-based speech synthesizer to the task of articulatory inversion, based on the successful text-to-speech WaveNet \cite{DBLP:journals/corr/OordDZSVGKSK16} system and its derivatives. WaveNet introduced a novel approach to speech synthesis based on point-to-point prediction of the raw audio signals. WaveNet takes audio waveform inputs and models acoustic information with a deep generative architecture composed of stacked dilated causal convolutional layers that model point by point conditional probabilities with a wide receptive field. Inspired by the success of WaveNet architectures in different speech synthesis tasks \cite{DBLP:journals/corr/OordDZSVGKSK16,tanaka2019wavecyclegan2,prenger2019waveglow,shen2018natural,
wang2017tacotron,hayashi2019espnet,paine2016fast,maiti2019parametric,kastner2019representation,
tamamori2017speaker,oord2017parallel,govalkar2019comparison}, we hypothesize that we can adapt this stacked dilated convolutional layer approach to modeling articulatory waveforms to improve the accuracy of articulatory inversion. 
The remaining sections of the paper are organized as follows: Section \ref{Background-Related-work} is a review of the technical background of our current GMM-HMM approach, comparative A2AI deep learning approaches, and the WaveNet architecture; Section \ref{Dataset} describes our dataset, the ElectroMagnetic Articulography corpus of Mandarin Accented English (EMA-MAE) ; Section \ref{Feature Description} introduces the acoustic and articulatory feature sets and pre-processing elements; Section \ref{Model Architecture } describes the proposed model architecture; Section \ref{Experiments and Results } describes the experimental methodology and results; and  Section \ref{Conclusion} summarizes the findings and presents the overall conclusions.  
\subsection{Background and Related work} \label{Background-Related-work}
\subsubsection{Baseline GMM-HMM approach for A2AI}
Our previous framework \cite{bozorg2018comparing} consisted of parallel acoustic and articulatory HMMs, with dynamic smoothing to account for the presence of discrete rather than continuous state variables. In the training phase, parallel acoustic-articulatory data was trained separately for each individual speaker. In the inversion stage, the test speech was input to the trained acoustic HMMs to derive an optimal HMM state alignment, and then the corresponding aligned articulatory HMMs were used to recover the articulatory trajectory. Once the alignment of articulatory states is complete, the recovery algorithm estimates a smooth articulatory trajectory from the HMM. Results with this method give an average correlation between actual measured and estimated trajectories of 0.61 and an average Root Mean-Square-Error (RMSE) of 2.83 mm.
\subsubsection{Deep Learning Architecture for A2AI}
There has been substantial recent research on deep learning approaches to mapping between acoustic waveforms and articulatory trajectories. For instance, Sivarman et al. \cite{sivaraman2017analysis} applied Artificial Neural Networks (ANN), Cail et al. \cite{cai2018dku} and Seneviratne et al. \cite{seneviratne2019multi} deployed a Deep Neural Network (DNN) architecture, Tobing et al. \cite{8282219} used a Latent Trajectory DNN, and Uria et al. \cite{uria2012deep}  utilized DNN and a deep trajectory-Mixture Density Network (MDN) for estimating articulatory trajectories from acoustic signals. Illa and Ghosh \cite{illa2020impact,illa2017comparative,illa2020closed,illa2019representation} have proposed two different DNN approaches \cite{illa2020impact,illa2017comparative}, Bidirectional Long-Short Term Memory (BLSTM) \cite{illa2020closed} and Convolutional Neural Network (CNN) layer cascaded to the BLSTM network \cite{illa2019representation} for speaker dependent A2AI. Mannem et al. \cite{mannem2019acoustic} used a convolutional dense neural network, Richmond et.al \cite{richmond2007trajectory,richmond2006trajectory} used an MDN architecture and augmented static articulatory features with dynamic features. To capture dependencies between articulatory trajectories and corresponded past, current and future acoustic features Liu et al. \cite{7178812} implemented BLSTM and deep recurrent MDN. Xie et al. \cite{8706709} investigated different architectures such as DNN, Recurrent Neural Network (RNN), MDN, Time Delay DNN-MDN, RNN-MDN and RNN-MDN BLSTM, Biasutto et al. \cite{biasutto2018phoneme} applied bidirectional gated RNN and Maud et al. \cite{maud2019independent} make use of the BLSTM neural network with an additional convolutional layer, which acts as a low pass filter after the readout layer for A2AI. 
The best of these approaches have RMSE in the range of 0.6-0.8 mm \cite{uria2012deep,7178812,8706709,biasutto2018phoneme} on the MNGU0 \cite{richmond2011announcing} dataset, and correlations of  0.8-0.9 \cite{illa2020impact,illa2017comparative,illa2019representation,richmond2007trajectory,7178812,8706709,biasutto2018phoneme} have been demonstrated on the  MOCHA\cite{Wrench00amultichannel} and MNGU0 datasets. It can be very difficult to compare A2AI results across different datasets, especially using the RMSE metric method. RMSE is highly speaker dependent and therefore varies across different speakers and corpuses. For example, several approaches have demonstrated RMSE of less than 1mm on the MNGU0 corpus, but the same methods have an RMSE of around 1.5-2 mm for the MOCHA corpus \cite{illa2019representation,richmond2007trajectory,tobing2017deep}. Seneviratne et al. \cite{seneviratne2019multi} compared A2AI with a DNN architecture across different corpuses and training sets. They reported correlation results of several different articulators within and across corpora. For example, the correlations of the actual and predicted trajectories of the Lip Protrusion (LP) articulatory feature were 0.60, 0.76 and 0.62 for XRMB \cite{macneilage1982speech}, EMA-IEEE \cite{institute1969ieee} and MOCHA-TIMIT datasets, respectively, when trained and tested on data from the same corpus. For multi-corpus training, the correlation results were 0.57, 0.75 and 0.66 for these same three corpuses. 
Although both RMSE and correlation vary substantially by speaker and dataset, correlation varies much less and therefore tends to be a more consistent and stable metric for evaluation. The best and newest reported methods based on deep learning architectures have achieved between 0.8-0.85 correlation on average \cite{illa2020impact,illa2017comparative,richmond2007trajectory,7178812,biasutto2018phoneme}. In comparison,  traditional methods such as GMM and GMM-HMM are substantially lower, between 0.55-0.65 \cite{ji2014speaker,toda2008statistical,zhang2008acoustic,bozorg2018comparing}.
 
The EMA-MAE dataset used here, with 39 speakers, has significantly more speaker variability compared to other datasets like MNGU0 and MOCHA which contain records from just one or two speakers.
\subsubsection{WaveNet}
Googleâ€™s WaveNet architecture \cite{DBLP:journals/corr/OordDZSVGKSK16} is a novel approach to the problem of speech waveform synthesis that has significantly improved intelligibility for text-to-speech applications. The dilated causal convolution architecture at WaveNet \cite{DBLP:journals/corr/OordDZSVGKSK16} addresses the problem of long-range temporal tracking of raw audio data for speech synthesis.
The new proposed A2AI model, Articulatory-WaveNet, like many other Modified-WaveNet approaches, deploys architectures conditioned on acoustic characteristics like Mel-Spectrograms instead of linguistic features \cite{DBLP:journals/corr/OordDZSVGKSK16} . For instance, Kastner \cite{kastner2019representation}, Tacotron \cite{shen2018natural} and Tacotron 2 \cite{wang2017tacotron} used attention based RNN to extract Mel-Spectrograms features from input information  for conditioning their WaveNet model. Conditioned WaveNet with acoustic features have been also used in  \cite{tanaka2019wavecyclegan2,prenger2019waveglow,maiti2019parametric,tamamori2017speaker
,spratley2019unified}.
Recently, many studies have been conducted to improve the WaveNet architecture and make the generation and synthesis of samples faster. For example, the Probability Density Distillation (PDD) method \cite{oord2017parallel} combines two strategies of Inverse Autoregressive Flows (IAF) and WaveNet to make the system compatible with real-time processing and parallel computing. The PDD utilizes a teacher trained WaveNet to transform the knowledge and train the parallel feed forward IAF (student) network. This system is much faster than the vanilla WaveNet, therefore it can be used for variant languages and multiple speakers. To eliminate unnecessary convolutional operations, Paine et al. \cite{paine2016fast}  proposed a new method named Fast-WaveNet. This framework caches previous computations instead of recomputing them from scratch to predict the new samples. Compared to the naive WaveNet, Fast-WaveNet reduces the complexity of the operation from $O(2^L)$ to $O(L)$, which $L$ represents number of layers in the neural network. In the work presented here we have used this Fast-WaveNet approach with our Articulatory-WaveNet framework to generate articulatory trajectories faster.
\section{Dataset}\label{Dataset}
The EMA corpus of Mandarin Accented English (EMA-MAE) \cite{ji2014electromagnetic}  has been used for this study. This dataset is one of the largest of its kind, with 39 total speakers. Speakers consist of an L1 group of 10 males and 10 female native English speakers (upper Midwest accent Standard American English) and an L2 group of 10 males and 9 female native Mandarin speakers (two dialect regions, including Beijing accent and Shanghai accent Modern Standard Mandarin). For each individual speaker, about 45 minutes of acoustic and articulatory data were collected, including word, sentence, and paragraph-level speech samples. A bite-plate calibration technique was used to transform the data to a standardized articulatory coordinate system. The resulting co-ordinate system for the data is based on the midsagittal plane and maxillary occlusal plane of the speaker with an origin at the lowest mid-point of the upper incisors.
For each speaker, a sensor-tipped wand was used to collect a map of the hard palate. This information allows for the calculation of articulatory variables representing vocal-tract distances in addition to raw sensor positions.
Sensors were placed at the midsagittal plane locations of the lower Middle Incisor (MI), Lower Lip (LL), Upper Lip (UL), Tongue Dorsum (TD), and Tongue Tip (TA). In addition, there were two lateral sensors, one at the Left Corner (LC) of the mouth to help indicate lip corner and one in the Left central midpoint (LT) of the Tongue body to help indicate lateral tongue curvature.
\section{Feature Description}\label{Feature Description}
The articulatory feature set used for our A2AI experiments with the EMA-MAE corpus consists of 6 tongue-related features, 3 lip-related features, and a jaw feature.  The tongue features include the 3 horizontal distances to the Tip, dorsum and lateral sensors and the 3 vertical distances between the sensors and the hard palate. Lip features include lip protrusion, lip separation, and lateral distance to the corner lip sensor, which is indicative of lip rounding. Table \ref{table_VT} represents the articulatory feature set applied for evaluating Articulatory-WaveNet.  
%%%%%%%%%%%%%%%%%%%%%%%%%%%%%%%%%%%%%%%%%%%%%%%
\begin{table}[!h]
%% increase table row spacing, adjust to taste
\renewcommand{\arraystretch}{1.3}
\caption{Vocal Tract Features}
\label{table_VT}
\centering
%% Some packages, such as MDW tools, offer better commands for making tables
%% than the plain LaTeX2e tabular which is used here.
\begin{tabular}{|c|}
\hline
Tongue Dorsum Horizontal Position \\
\hline
Tongue Dorsum Vertical Height to Hard Palate \\
\hline
Lateral Tongue Horizontal Position \\
\hline
Lateral Tongue Vertical Height to Hard Palate \\
\hline
Tongue Tip Horizontal Position \\
\hline
Tongue Tip Vertical Height to Hard Palate \\
\hline
Horizontal Lip Protrusion \\
\hline
Vertical Lip Separation \\
\hline
Lateral Lip Distance (Lip Corner Sensor) \\
\hline
Vertical Middle Incisor (Jaw) \\
\hline
\end{tabular}
\end{table}
%%%%%%%%%%%%%%%%%%%%%%%%%%%%%%%%%%%%%%%%%%%%%%%%
For the acoustic data, Mel-Spectrograms features are used \cite{deller2000discrete}. 
Articulatory features were calculated point-by-point on the 400Hz EMA data, then downsampled by a factor of 4 to give one feature every 10ms.  
\section{ Model Architecture }\label{Model Architecture }
%%%%%%%%%%%%%%%%%%%%%%%%%%%%%%%%%%%%%%%%%%%%%%%%%%%%%%
\begin{figure*}[!t]
\centering
\includegraphics[width=6.5in,height=3.5in]{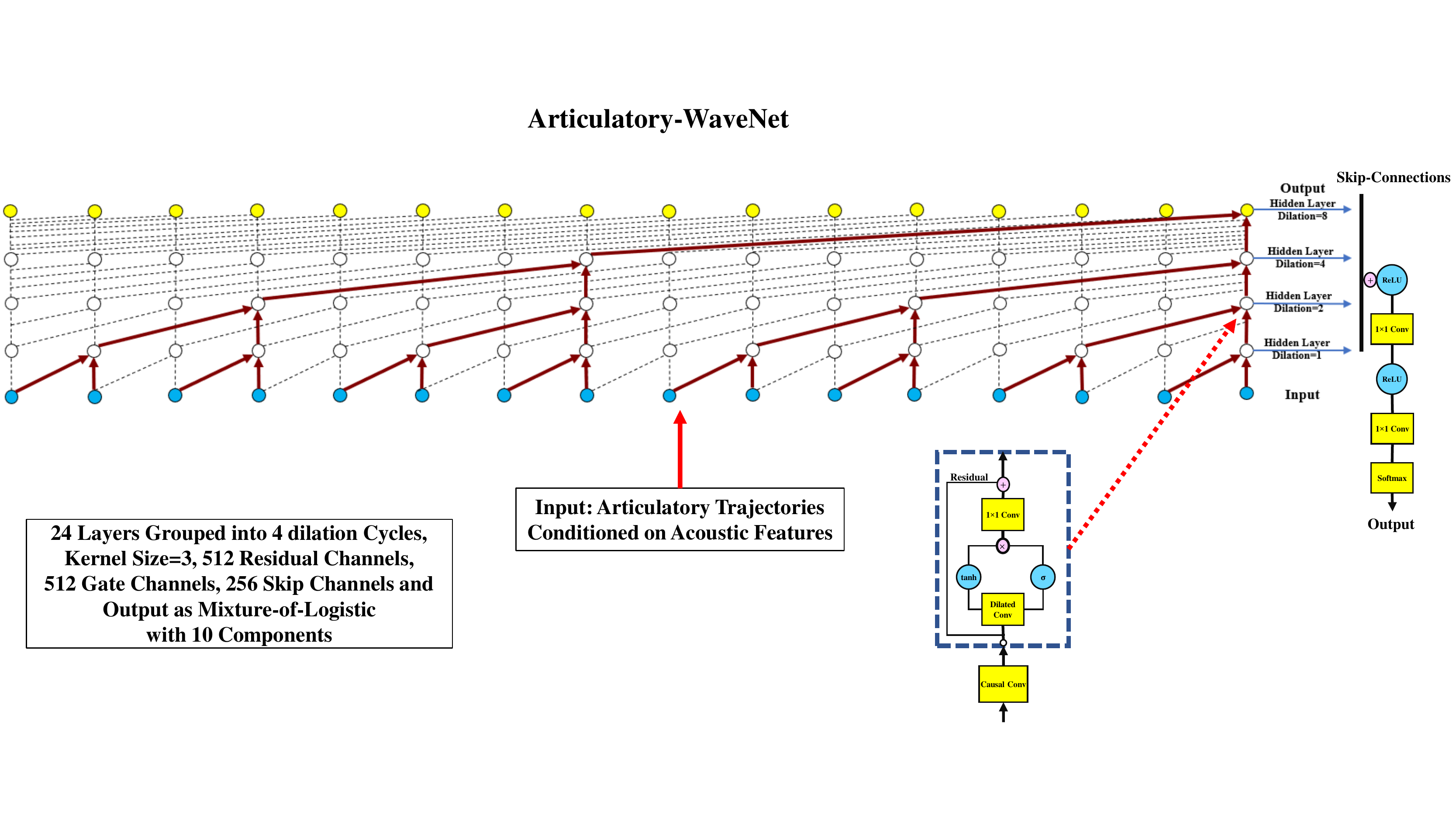}
% where an .eps filename suffix will be assumed under latex, 
% and a .pdf suffix will be assumed for pdflatex; or what has been declared
% via \DeclareGraphicsExtensions.
\caption{Visualization of Articulatory-WaveNet,stacked causal convolutional layers, with overview of the residual block and overall architecture.}
\label{fig_arc}
\end{figure*}
%%%%%%%%%%%%%%%%%%%%%%%%%%%%%%%%%%%%%%%%%%%%%%%%%%%%%%%%%%%5
The Articulatory-WaveNet architecture introduced here is a fully probabilistic and autoregressive model that generates a time-series articulatory trajectory by using the causal conditional predictive distribution of samples. This architecture utilizes stacked convolutional layers to model the conditional probability distribution. The product of the sequential conditional probabilities over time is represented as a model of the joint probability of a time-series signal. The occurrence of each sample  from articulatory trajectory, $x$, is  conditioned on all previous samples $(x_1,..,x_{t-1})$ \cite{DBLP:journals/corr/OordDZSVGKSK16,oord2017parallel} :
\begin{equation}\label{prb_causal}
 P(x)=\displaystyle\prod_{t=1}^{T} p(x_t \mid x_1,...,x_{t-1})
\end{equation}
By using the probabilistic chain-rule and product of the conditional distributions, the
autoregressive Articulatory-WaveNet network models the joint distribution of high-dimensional data like articulatory trajectories.
\subsection{Dilated Causal Convolutions}
The causal or masked convolutional layers play the main role in the autoregressive convolutional Articulatory-WaveNet model. By using this framework all dependencies on future events or samples are omitted and all $P(x_t|x<t)$  can be generated in one forward pass. The prediction of the sample   only relies on the previous events $(x_1,..,x_{t-1})$ as it represented in equation \ref{prb_causal} \cite{DBLP:journals/corr/OordDZSVGKSK16,oord2017parallel} . This can be implemented on kinematic time-series trajectories by shifting the results of regular convolution for a few time steps.
In addition, dilated causal convolutions have enabled Articulatory-WaveNet to deal with the high dimensional articulatory trajectories (400 samples per second). This technique not only allows the network to grow its receptive field exponentially with depth, but it also supports the standard causal convolution layers for modeling the long-term dependencies with a sufficiently wide receptive field.
 
The size of the receptive field of the causal convolutional neural network is computed by following equation \cite{DBLP:journals/corr/OordDZSVGKSK16}:
\begin{equation}
\textrm{Receptive Field = Number of Layers + Filter Length - 1}
\end{equation}
The larger receptive field requires either more layers or a larger filter. However, using dilated convolutional layers provides a vast receptive field by dilating the original filter with zeros. This type of architecture not only enlarges the receptive field but also keeps the computational costs and input resolution at the same value.  The simple CNN model is a type of dilated convolutional architecture with dilation set to the 1 \cite{DBLP:journals/corr/OordDZSVGKSK16,oord2017parallel}.
Fig.\ref{fig_arc}. the Articulatory-WaveNet with a stack of causal convolutional layers.
In this architecture, the gates are the nonlinear activation units for modeling the time-series signal. Gate activation for the input $x$ and gate output $z$ is computed by:
\begin{equation}
\rm{z=\tanh(w_{f,k}*x)\odot\sigma(w_{g,k}*x)}
\end{equation}
where $*$ represents the convolutional operator, $\odot$ is an element-wise multiplication operator, $\sigma(.)$ denotes a logistic sigmoid function, $k$ is the layer index, $f$ and $g$ are filter and gate indices, respectively, and $w$ is the convolutional filter weight matrix \cite{DBLP:journals/corr/OordDZSVGKSK16}. 
In order to build a deeper architecture and raise the speed of convergence, residual and parametrized skip connections have been also used in Articulatory-WaveNet. These operators are repeated in each stack.
\subsection{Conditioning }
Articulatory-WaveNet has the ability to model the sequence of articulatory trajectories which have been conditioned on the sequence of additional time-series acoustic features. By conditioning the network, the predicted articulatory trajectories will be based on the acoustic information \cite{DBLP:journals/corr/OordDZSVGKSK16}. The conditional probability distribution is represented by:
 
 \begin{equation}
\rm{p(x|h_t)=\prod_{t=1}^{T} p(x_t \mid x_1,...,x_{t-1},h_t)}
\end{equation}
 
where $h_t$ represents the conditioning Mel-Spectrograms. Including the additional conditional input, the activation unit in equation \ref{prb_causal} turns into:
\begin{equation}
\rm{z=\tanh(w_{f,k}*x+V_{f,k}*h(t))\odot\sigma(w_{g,k}*x+V_{g,k}*h(t))}
\end{equation}
\section{Experiments and Results }\label{Experiments and Results }
We evaluated the performance of the Articulatory-WaveNet A2AI framework using the EMA-MAE corpus. 
 
\subsection{Data Preparation}
Mel-Spectrograms are extracted through a Hanning-windowed Short-Time Fourier Transform with 38.7 ms frame size and 9.7 ms frame hop.  Log dynamic range compression is implemented using a 80 channel mel filter bank spanning the range of 125 Hz to 7.6 kHz.
The articulatory features are the 10 static features described in section \ref{Feature Description}. The WaveNet architecture has $\tanh$ activation functions which output in the $[-1 \ 1]$ range, so the articulatory inputs have been scaled to this range using global dynamic range normalization: 
\begin{eqnarray}
\textrm{(Scaled Articulatory Feature)}_i = \\\newline \nonumber
2\left(\frac{\textrm{Articulatory Feature} - \textrm{Min}_i}{Max_i-Min_i}-1\right)
\end{eqnarray}
The dynamic range normalization is unique to each speaker and articulatory variable, with $Max_i$  and $Min_i$ representing the overall maximum and minimum of all articulatory trajectories for speaker $i$. This structure allows for easy conversion of predicted trajectories to the original feature space. 
For this set of experiments, the utterances from EMA-MAE were applied to prepare training and test sets for evaluating the performance of the Articulatory-WaveNet A2AI framework. For the training set, 4000 utterances were randomly selected across all the speakers (102-103 utterances per speaker), while for the test set another 580 utterances (14-15 utterances per speaker) were randomly selected. The training and test sets were selected separately, and they include sentence and word speech samples.  
%%%%%%%%%%%%%%%%%%%%%%%%%%%%%%%%%%%%%%%%%%%%5
\begin{table*}[t!]
%% increase table row spacing, adjust to taste
\renewcommand{\arraystretch}{1.3}
%\setlength{\columnsep}{2}
% if using array.sty, it might be a good idea to tweak the value of
\caption{Performance Comparision of the Articulatory-WaveNet and HMM-GMM}
\label{table_EVAL}
\centering
\begin{tabular}{| c | c | c | c | c | c | c | }
\hline
\textbf{ArticulatoryTrajectories} & \multicolumn{3}{ c |}{\textbf{CORRELATION}} & \multicolumn{3}{ c |}{\textbf{RMSE (Millimeters)}}\\ 
\cline{2-7}
& \textbf{HMM-GMM} & \textbf{ART-WN} & \textbf{\%increase} & \textbf{HMM-GMM}& \textbf{ART-WN} & \textbf{\%decrease}  \\
\hline
Horizontal Tongue Dorsum (VT1) & 0.59 &0.84& 42.3 &3.41 &1.14 &66.5  \\ \hline
Tongue Dorsum
Vertical Height to Hard Palate (VT2) &  0.64 &0.82& 28.1& 3.44& 1.24 &63.9\\ \hline
Horizontal Lateral Tongue (VT3)&  0.61 &0.83 &36.1 &2.62& 1.40& 46.5 \\ \hline
Lateral Tongue
Vertical Height to Hard Palate (VT4) & 0.66& 0.82& 24.2& 2.35 &1.29& 45.1\\ \hline
Horizontal Tongue Tip (VT5) & 0.62 &0.83 &33.9& 3.28& 1.62& 50.6 \\ \hline
Tongue Tip 
Vertical Height to Hard Palate (VT6)&  0.65& 0.82 & 26.2& 3.37& 1.66& 50.7 \\ \hline
Horizontal Lip Protrusion (VT7)&  0.55 &0.82& 49.1& 3.37& 0.26& 92.2\\ \hline
Vertical Lip Separation (VT8)& 0.61 &0.84& 37.7& 3.05& 1.65& 45.9 \\ \hline
Lateral Lip Corner (VT9)&  0.50& 0.82& 64.0& 0.83& 0.18& 78.3\\ \hline
Vertical Middle Incisor (Jaw) (VT10)&0.67 &0.81& 20.9& 1.99& 2.08& -4.3 \\ \hline
MEAN &0.61 &0.83 &36.1& 2.83& 1.25& 55.8\\ \hline
\end{tabular}
\end{table*}
\begin{table*}[h]
%% increase table row spacing, adjust to taste
\renewcommand{\arraystretch}{1.3}
% if using array.sty, it might be a good idea to tweak the value of
% \extrarowheight as needed to properly center the text within the cells
\caption{Performance Comparision of the Articulatory-WaveNet for the different L1/L2 and Male/Female subgroups.}
\label{table_subgroup}
\centering
%% Some packages, such as MDW tools, offer better commands for making tables
%% than the plain LaTeX2e tabular which is used here.
\begin{tabular}{| c |c |c |}
\hline
\shortstack{\\Gender }& \shortstack{\\RMSE (Millimeter) Results for Articulatory Trajectories\\
\  VT1 \ \ \ VT2 \ \ \ VT3 \ \ \ VT4 \ \ \ VT5\ \ \ VT6\ \ \ VT7\ \ \ VT8\ \ \ VT9\ \ \ VT10 }& Average \\
\hline
Male & \ 1.24 \ \ \ 1.50 \ \ \ 1.56 \ \ \ 1.36 \ \ \ 1.81 \ \ \ 1.88 \ \ \ 0.23 \ \ \ 1.67 \ \ \ 0.18 \ \ \ 1.84  &1.33\\
\hline
Female& \ 1.04 \ \ \ 0.97 \ \ \ 1.23 \ \ \ 1.21 \ \ \ 1.42 \ \ \ 1.43 \ \ \ 0.30 \ \ \ 1.63 \ \ \ 0.19 \ \ \ 2.34  &1.18\\
\hline
{Gender}& \shortstack{\\Correlation Results for Articulatory Trajectories\\
\  VT1 \ \ \ VT2 \ \ \ VT3 \ \ \ VT4 \ \ \ VT5\ \ \ VT6\ \ \ VT7\ \ \ VT8\ \ \ VT9\ \ \ VT10 }& Average \\
\hline
Male & \ 0.84 \ \ \ 0.81 \ \ \ 0.82 \ \ \ 0.82 \ \ \ 0.81 \ \ \ 0.80 \ \ \ 0.83 \ \ \ 0.82 \ \ \ 0.80 \ \ \ 0.80 & 0.82\\
\hline
Female & \ 0.83 \ \ \ 0.82 \ \ \ 0.82 \ \ \ 0.81 \ \ \ 0.82 \ \ \ 0.81 \ \ \ 0.82  \ \ \ 0.83  \ \ \ 0.82 \ \ \ 0.80 & 0.82\\
\hline
{L1/L2}&\shortstack{\\RMSE (Millimeter) Results for Articulatory Trajectories\\
\  VT1 \ \ \ VT2 \ \ \ VT3 \ \ \ VT4 \ \ \ VT5\ \ \ VT6\ \ \ VT7\ \ \ VT8\ \ \ VT9\ \ \ VT10 }& Average \\
\hline
MN & 1.16  \ \ \ 1.28  \ \ \ 1.91  \ \ \ 1.62  \ \ \ 2.35  \ \ \ 1.71  \ \ \ 0.33  \ \ \ 1.81  \ \ \ 0.16  \ \ \ 2.06 & 1.44\\
\hline
EN & 1.13  \ \ \ 1.21  \ \ \ 0.91  \ \ \ 0.98  \ \ \ 0.93  \ \ \ 1.62  \ \ \ 0.20  \ \ \ 1.50  \ \ \ 0.20  \ \ \ 2.11 &1.08  \\
\hline
{L1/L2}& \shortstack{\\Correlation Results for Articulatory Trajectories\\
\  VT1 \ \ \ VT2 \ \ \ VT3 \ \ \ VT4 \ \ \ VT5\ \ \ VT6\ \ \ VT7\ \ \ VT8\ \ \ VT9\ \ \ VT10 }& Average \\
\hline
MN & 0.83  \ \ \ 0.82  \ \ \ 0.82  \ \ \ 0.81  \ \ \ 0.81  \ \ \ 0.80  \ \ \ 0.83  \ \ \ 0.84 \ \ \ 0.80  \ \ \ 0.81 & 0.82\\
\hline
{EN} & 0.84  \ \ \ 0.81  \ \ \ 0.82  \ \ \ 0.82  \ \ \ 0.82  \ \ \ 0.81  \ \ \ 0.83  \ \ \ 0.81  \ \ \ 0.81  \ \ \ 0.79 & 0.82 \\
\hline
\end{tabular}
\end{table*}
%%%%%%%%%%%%%%%%%%%%%%%%%%%%%%%%%%%%%%%%%%%%%%%%%%%%%%%%%%%%%%%%%
\subsection{Training and Synthesizing Articulatory Trajectories}
To speed up the synthesizing process, Articulatory-WaveNet uses the Fast-WaveNet Generation Algorithm \cite{paine2016fast}. Fast WaveNet caches previously computed information from the overlapping network states, called recurrent states, to eliminate redundant convolutions. To implement this, the network computes a new output sample using the caching information from the recurrent states. This is a significant computational improvement over WaveNet which re-computes all states at each time step. 
Articulatory-WaveNet has 24 layers with 4 dilation stacks. The dilation rate increases by a factor of 2 in every layer at each stack. This starts with no dilation (rate 1) and reaches a maximum dilation of 512. For this experiment we considered 4 stacks: $1,2, 4, ..., 512,1, 2, 4, ..., 512, 1, 2, 4, ..., 512,1, 2,4, ..., 512$. The stacking grows the receptive field size and increases the capacity of the network.  The kernel size of the causal dilated convolutions is 3, with 512 units in the gating layers and residual connection channels and 256 hidden units at the skip connection channel and $1*1$ convolution before the output layer. The output is modeled as a mixture of 10 logistic components for higher quality.
To compute the logistic mixture distribution, the Articulatory-WaveNet stack output is passed through a ReLU activation followed by a linear projection to predict parameters $\theta=\{\textrm{ Mean }\mu_i ,\textrm{ Log Scale } S_i,\textrm{ Mixture Weight } \pi_i\}$  for each mixture component. The loss is computed as the negative log-likelihood of the ground truth sample. The likelihood of sample $x_t$ is:
\begin{equation}
\rm{P(x_t|\theta,h_t)=\sum_{i=1}^{k=10}\pi_i[\sigma(\frac{\tilde{x}_{ti}+0.5}{S_i})-\sigma(\frac{\tilde{x}_{ti}-0.5}{S_i})]}
\end{equation}
where $\tilde{x}_{ti}=x_t-\mu_i$ and $P(x_t|\theta,h_t)$ is the probability density function of the articulatory trajectory conditioned on mel-spectrogram $h_t$.
The Articulatory-WaveNet network was trained for 20,000 epochs using the ADAM optimizer. There are 8 mini-batches with each minibatch containing a maximum of 8000 timesteps (roughly 302ms). 
%%%%%%%%%%%%%%%%%%%%%%%%%%%%%%%%%%%%%%%%%%%%%%%%%%%
\begin{figure*}[!ht]
\centering
\includegraphics[width=3in,height=2in]{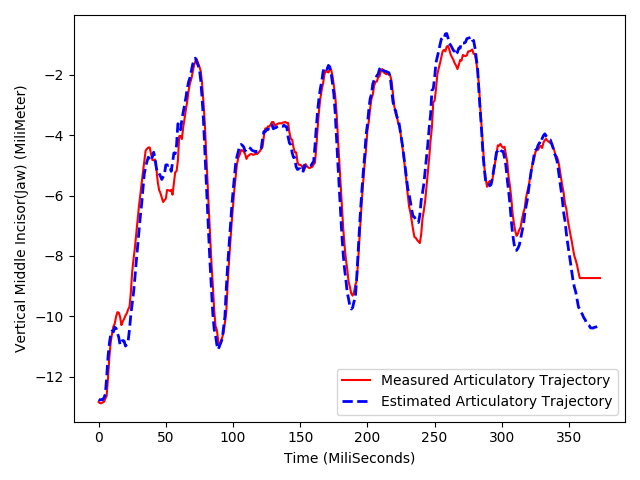}
\includegraphics[width=3in,height=2in]{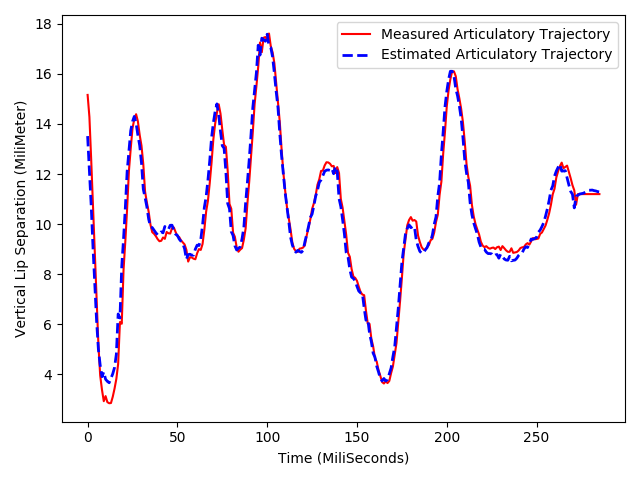}
\includegraphics[width=3in,height=2in]{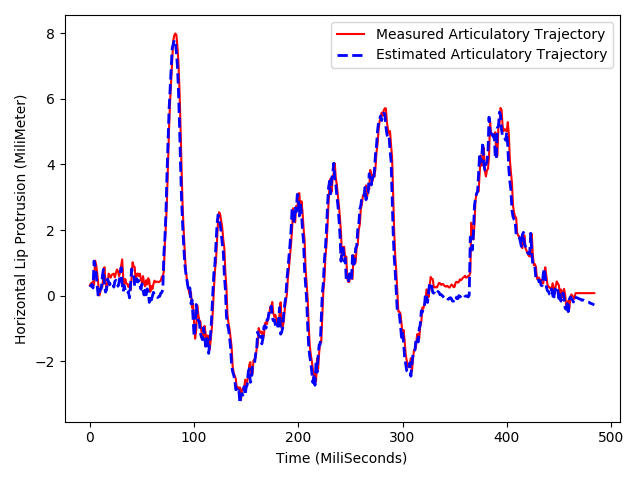}
\includegraphics[width=3in,height=2in]{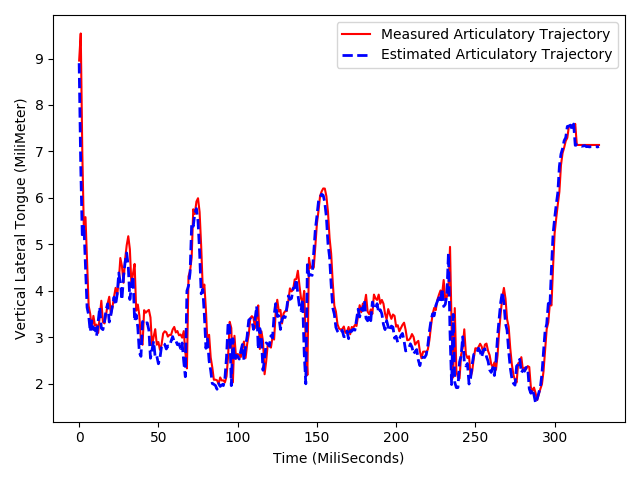}
\includegraphics[width=3in,height=2in]{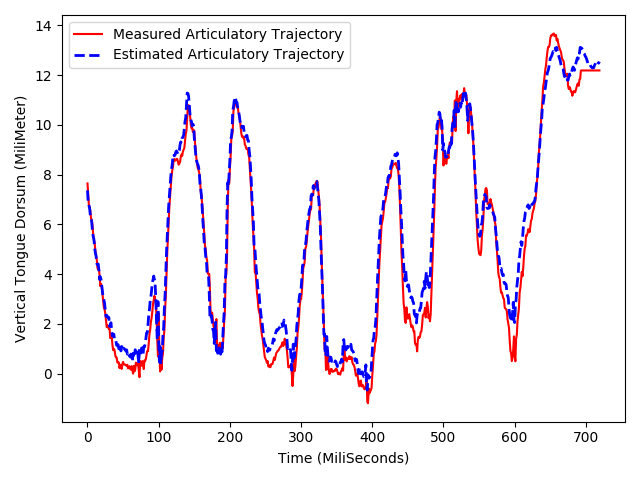}
% where an .eps filename suffix will be assumed under latex, 
% and a .pdf suffix will be assumed for pdflatex; or what has been declared
% via \DeclareGraphicsExtensions.
\caption{Trajectories of selected articulatory features from a typical test sentence utterances. The plots show the trajectories that have been estimated by Articulatory-WaveNet alonside the target actual articulatory trajectories.}
\label{fig_TRJ}
\end{figure*}
%%%%%%%%%%%%%%%%%%%%%%%%%%%%%%%%%%%%%%%%%%%%%
\subsection{A2AI Results}
Evaluation metrics for A2AI include RMSE and correlation coefficient. The RMSE is computed by the following equation:
\begin{equation}
\rm{E_{rms}=\sqrt{\frac{1}{m}\sum_{i=1}^{m}(f(x_i)-y_i)^2}}
\end{equation}
Where $y$ are the known values, $f(x)$ is the estimated output trajectory, and $m$ is the number of test files.
Results are also evaluated using a correlation metric between actual and estimated trajectories:
\begin{equation}
\rm corr=\frac{\displaystyle\sum_{i=1}^{m}(f(x_i)-\overline{f(x)})(y_i-\bar{y})}{{\displaystyle\sum_{i=1}^{m}}(f(x_i)-\overline{f(x)})^2 \ {\displaystyle\sum_{i=1}^{m}}(y_i-\bar{y})^2}
\end{equation}
where $y$ are the known values, $f(x)$ is the estimated output, $m$ is the number of test files and $\overline{f(x)}$ , $\bar{y}$ are the utterance-level means of the estimated and actual trajectories. 
Table \ref{table_EVAL} shows the overall RMSE and correlation results of each individual articulatory feature averaged across all 39 speakers in the EMAMAE dataset.
Overall, the Articulatory-WaveNet improved correlation from 0.61 to 0.83 (36\% increase) and decreased RMSE from 2.83mm to 1.25mm (56\% decrease) over the baseline GMM-HMM system, averaged across all speakers (both native English and native Mandarin) and articulatory features.
 
Looking at RMSE specifically, the most significant improvements are for the horizontal Lip Protrusion, reduced from 3.37mm to 0.26mm (92.28), lateral Lip Corner, reduced from 0.83mm to 0.18mm (78.31), vertical and horizontal Tongue Dorsum, reduced from 3.44 and 3.41mm to 1.44 and 1.24mm (66\%  and 63.95\% decrease), and vertical and horizontal Tongue Tip, reduced from 3.37 and 3.28mm to 1.66 and 1.62mm (51\% decrease).
The average RMSE for tracking the vocal tract height at the three tongue sensors, key variables for capturing physiological characteristics of tongue motion, is 1.39mm, down from 3.05mm for the baseline method.  Speaker horizontal tongue sensor positions have an average RMSE of 1.38mm, down from 3.10mm. Vertical lip separation had an RMSE of 1.65mm, down from 3.05mm. Horizontal lip protrusion and Lateral lip distance both show slightly lower RMSEs 0.26mm and 0.18mm, down from 3.37mm and 0.83mm respectively. Middle incisor (jaw) sensor show slightly higher RMSE 2.08mm compared to baseline 1.99mm, which is interesting since it showed improved correlation.

Correlation results show consistent improvement across all features, with all 10 of the articulatory feature trajectories having correlations above 80\%, ranging from 81\% to 84\%.
 Fig.\ref{fig_TRJ}. shows the measured EMA and estimated articulatory movements for a selection of speakers and articulatory features, for visualization of the results.
 
In addition, we also compared the performance of the Articulatory-WaveNet for different subgroups of speakers. The results show consistency of performance of the proposed architecture for predicting articulatory features from acoustic features across different subgroups of speakers including Native speakers (L1), second language speakers (L2), and Male and Female speakers. Table \ref{table_subgroup} show the RMSE and Correlation results for these different groups of speakers.
 
Results indicate that the correlation is consistent across different type of L1, L2, male and female groups of speakers and it remains around 82\%. However, RMSE results differ across speakers. 
Comparison of RMSE results for L1 and L2 speakers show that for L1 results are more accurate, with L2 speakers having 0.36 mm higher RMSE. This is consistent with what might be expected for L1 vs. L2 speaker groups in terms of pronunciation and articulatory consistency.

Looking at results for L1 English speakers in particular, for several articulatory features including lips, tongue and incisor Articulatory-WaveNet show improvement compared to the best reported approaches. The average RMSE from Latent Trajectory DNN \cite{tobing2017deep} approach for the vertical tongue (tip, body and dorsum) is around 1.80mm while for Articulatory-WaveNet the vertical tongue (tip, lateral and dorsum) RMSE for English speakers is 1.27mm. The best reported results with CNN+BLSTM approach in \cite{illa2019representation} for 12 articulatory features including lip, jaw and tongue is reported around 0.84 correlation and 1.4mm RMSE. The Articulatory-WaveNet approach for English speakers has a similar correlation, at 0.82, but an RMSE of only 1.08mm.

In addition, female speakers have slightly better results for  RMSE compared to the male speakers, 0.15 mm lower. 
 
\section{Conclusion}\label{Conclusion}
In this paper, we have proposed the Articulatory WaveNet architecture for A2AI. The model has been tested on the EMA-MAE corpus and shows significant improvement for RMSE and Correlation compared to the baseline GMM-HMM system, with correlations above 80\% for all articulatory trajectory estimates and an average RMSE of 1.25mm across both L1 and L2 speaker groups. Within native English speakers, average RMSE across the set of ten articulatory features for the proposed method is 1.08mm, demonstrating state-of-the-art results on the A2AI task.
\bibliography{Wavenet1}
\bibliographystyle{IEEEtran}
\end{document}